\documentclass[prd,aps,12pt,nofootinbib,reprint]{revtex4-1} 
\usepackage{mathrsfs,amssymb,amsmath,amsxtra,amsfonts}
\usepackage{graphicx,color}
\usepackage{verbatim}
\usepackage[colorlinks]{hyperref}
\usepackage[utf8]{inputenc}

\begin{document}

\title{EPR-Bell realism as a part of logic}
\author{I. Schmelzer}
\noaffiliation

\begin{abstract}
We argue that for the proof of Bell's theorem no assumptions about realism or free will are necessary. The key formula 
\[E(AB|a,b) = \int A(a,b,\lambda)B(a,b,\lambda)\rho(\lambda) d\lambda\] 
follows from the logic of plausible reasoning (the objective Bayesian interpretation of probability theory) taken alone, without any further assumptions about realism. The space $\Lambda$, usually interpreted as some space of ``hidden variables'', can be constructed for an arbitrary ``field of discourse'' using Stone's theorem. The rejection of superdeterminism follows from logical independence -- the non-existence of information which suggest a dependence -- of the free decisions of the experimenters from everything else. 

To prove the Bell inequality it is, then, sufficient to reduce this to
\[E(AB|a,b) = \int A(a,\lambda)B(b,\lambda)\rho(\lambda) d\lambda.\] 
This follows for space-like separated measurements from Einstein causality alone. So, the consequence of the violation of Bell's inequality is that Einstein causality has been empirically falsified, without any loopholes left. 

We consider and reject the idea of weakening the logic of plausible reasoning could be used to create a loophole. Finally, we discuss what can be used to replace Einstein causality. While weak (signal) Einstein causality holds, it follows either from a notion of strong causality or from human inability to prepare states which violate quantum uncertainty, which makes it uninteresting for fundamental considerations. A notion of strong causality has either unacceptable causal loops or requires a hidden preferred frame. Given that a hidden preferred frame is completely compatible with modern physics, it is argued that this choice is preferable. 
\end{abstract}

\maketitle

\section{Introduction}

While there is a lot of discussion about the assumptions necessary to prove the Bell inequalities, there is a wide agreement that one needs, beyond Einstein causality (usually named ``locality''\footnote{We will not use ``locality'' because it is misleading. One can easily imagine theories with a preferred frame which are local in the common sense meaning, but with a maximal speed of information transfer greater than the speed of light. Despite being local in any reasonable sense, they would have to be named ``nonlocal'' if ``local'' is used for Einstein causality.}), some other non-trivial physical assumptions. Variants of proofs of Bell's inequality may use different assumptions, in particular ``realism'' (however specified), ``counter-factual definiteness'', or Reichenbach's principle of common cause \cite{Wiseman}. Moreover, one also has to exclude superdeterminism. 

The main thesis of this paper is that there is no need for any physical assumptions beyond Einstein causality. All what is necessary beyond Einstein causality is, in fact, already part of logic -- of the logic of plausible reasoning, usually named ``objective Bayesian interpretation'' of the rules of probability theory, as defined by Cox \cite{Cox} and Jaynes \cite{Jaynes}. The straightforward consequence is that, once the violation of Bell's inequalities for space-like separated measurements is accepted, Einstein causality has to be given up, as falsified by these experiments.

Beyond this main result, we discuss possible loopholes and consequences.

There is no way to avoid this conclusion beyond the questioning of logic, even if not of classical logic, but only in the less well-studied form of the logic of plausible reasoning. Given that defenders of Einstein causality have been ready to reject even such fundamental things as realism (as defined by the EPR criterion of reality) as well as classical causality (in any version containing Reichenbach's principle of common cause), one can expect that at least some physicists will consider even the rejection of logic as a reasonable alternative. Once there can be, even in principle, no logical arguments against the rejection of logic, we cannot prove they are wrong, so that we simply have to accept this possibility as a fact of life. Nonetheless, we give a short overview of previous attempts to question logic, because, even if they have failed to question logic itself, they were not completely worthless but have reached something, namely a better understanding of logic. Our suggestion is that questioning the logic of plausible reasoning has a chance to reach similar results too -- but that would not change the main result of this paper. It would leave the mathematics of the proof, and therefore the necessity to reject Einstein causality, unchanged. What one can reasonably hope for is a reinterpretation of these mathematics in common language, avoiding misinterpretations caused by the vagueness of human language, which we, as human beings, use to describe even precise mathematical results. 

For those who do not want to question the logic of plausible reasoning, the rejection of Einstein causality is unavoidable. What follows from this? The straightforward alternative would be the revival of a preferred frame, which would allow hidden quantum causal information transfer from past into future in some hidden absolute time coordinate. While this would be in contradiction with the standard interpretation of SR as well as GR, it would not be in conflict with observation. Indeed, in SR the Lorentz ether is simply an interpretation, even the original one. Moreover, viable alternatives with a preferred frame have been proposed for modern physics too, as for gravity \cite{gle}, as for the standard model of particle physics \cite{clm}. Thus, to accept a hidden preferred frame is at least not as problematic as usually assumed. 

Another alternative would be weakening causality, so that one can nonetheless preserve Einstein causality, even if only in a weaker sense. Here, we have a straightforward candidate worth to be discussed, namely weak or signal causality. Weak or signal Einstein causality could be violated only with a real transfer of information faster than light. Given the theorem that such information transfer is not possible using violations of the Bell inequalities, it seems reasonable to hope that this weaker notion could be preserved. And we, indeed, will not question this theorem, so, will not question signal Einstein causality. In this sense, it remains a way to preserve Einstein causality. 

Nonetheless, signal causality has no chance to have any fundamental relevance. Given the formulas which follow from the logic alone, there appears to be only one explanation why signal causality appears to be weaker than strong causality, namely that we are unable to prepare sufficiently localized states. If we would be able to prepare states with all really existing parameters fixed, we could send signals, but all what we can prepare are averages, quantum equilibrium states, and in these states all the information we would like to send is averaged out. So, it is the most trivial, straightforward excuse for our inability to send faster than light signals which appears to be sufficient. 

\section{What we need to prove Bell's inequality}

In the large amount of discussion following Bell's theorem, there has been no doubt (beyond the fringes) that Bell's inequality follows from eq. (2) of Bell's paper, which can be rewritten as the formula for the expectation value of the product $AB$ of the measurement results:
\begin{equation}\label{Bell-2}
E(AB|a,b) = \int A(a,\lambda)B(b,\lambda)\rho(\lambda) d\lambda.
\end{equation}
The part where Einstein causality is used can be also easily identified. If we can derive the formula
\begin{equation}\label{Bell-realistic}
E(AB|a,b) = \int_{\lambda\in\Lambda} A(a,b,\lambda)B(a,b,\lambda)\rho(\lambda) d\lambda,
\end{equation}
then we can derive \eqref{Bell-2} using Einstein causality, because Einstein causality tells us that the measurement result of Alice $A(a,b,\lambda)$ cannot depend on the decision $b$ of Bob what to measure, resp. for the measurement result of Bob $B(a,b,\lambda)$ depending on the decision $a$ of Alice. So, the question what we need beyond Einstein causality to derive Bell's inequality reduces to the question what we need to derive  \eqref{Bell-realistic}. Now, \eqref{Bell-realistic} is nothing but an application of the more general formula 
\begin{equation}\label{no-wishful-thinking}
E(o|a) = \int o(a,\lambda)\rho(\lambda) d\lambda,
\end{equation}
where $a\in A$ denotes all decisions of experimenters, and $o(a,\lambda)$ denotes some observable which is measured. What do we need to derive this formula? 

The straightforward way to derive it, which we will use here, is to obtain a general formula for the expectation values in Kolmogorovian probability theory, namely
\begin{equation}\label{Kolmogorovian}
E(f) = \int_{\omega\in\Omega} f(\omega) \rho(\omega) d \omega.
\end{equation}
Then we apply this formula to the general experimental situation where we distinguish between the parameters $a\in A$ which can be freely chosen by the experimenters, and all the other parameters $\lambda\in\Lambda$, so that instead of a general $\omega$ we consider now pairs $\omega=(a,\lambda)\in A\times\Lambda$. This would give 
\begin{equation}\label{wishful-thinking}
E(f) = \int f(a,\lambda)\rho(a,\lambda) da d\lambda.
\end{equation}
What remains is to get rid of the dependence of $\rho(a,\lambda)$ on $a$. That means, the free decisions of the experimenters should be independent of everything what defines the state of the system which is measured. This would give  
\begin{equation}\label{split}
E(f) = \int f(a,\lambda)\rho(a)\rho(\lambda) da d\lambda
\end{equation}
and as a consequence \eqref{no-wishful-thinking}. This step is the rejection of superdeterminism.  

So, what we have to do is to obtain formula \eqref{Kolmogorovian} of Kolmogorovian probability theory, and to justify the rejection of superdeterminism. 

\section{Kolmogorovian probability theory as the logic of plausible reasoning}

To justify the application of formulas of probability theory based on logic alone, we have to switch from the frequency interpretation of probability to the Bayesian interpretation of probability in its objective variant, as proposed by Cox \cite{Cox} and Jaynes \cite{Jaynes}. If we interpret \eqref{Kolmogorovian} as a formula of the frequency interpretation, we have no chance to derive it out of nothing but logic. Indeed, in the frequency interpretation any statements about probabilities are physical hypotheses about some frequencies, which, as any other empirical hypotheses, may be falsified by observation. Instead, in the objective Bayesian interpretation, the probabilities are degrees of plausibility of a statement given all the available information.  These degrees of plausibility have to follow logical rules of consistency. It appears that these rules of consistency, combined with a few simple conventions, are equivalent to the rules of probability theory.  

Unfortunately, the rules derived by Cox \cite{Cox} and Jaynes \cite{Jaynes} for the logic of plausible reasoning do not cover all formulas of Kolmogorovian probability theory. What Jaynes \cite{Jaynes} writes about differences between Kolmogorovian probability theory and the logic of plausible reasoning is not encouraging:
\begin{quote}
Our system of probability, however, differs conceptually from that of Kolmogorov in that we do not interpret propositions in terms of sets, but we do interpret probability distributions as carriers of incomplete information. \cite{Jaynes} p.50

The Kolmogorov system of probability (henceforth denoted by KSP) is a game played on a sample space $\Omega$ of elementary propositions $\omega_i$ (or ‘events’; it does not matter what we call them at this level of abstraction). \ldots For all practical purposes, then, our system will agree with KSP if we are applying it in the set-theory context. But in more general applications, although we have a field of discourse F and probability measure P on F with the same properties, we do not need, and do not always have, any set $\Omega$ of elementary propositions into which the elements of F can be resolved. \cite{Jaynes} p.651ff
\end{quote}

Fortunately, it appears that the set $\Omega$ can be constructed, always. Let's see. What is considered in the logic of plausible reasoning is a ``field of discourse'' $\mathcal{F}$, which is the set of all propositions considered. It is an abstract Boolean algebra, that means, we can apply the Boolean operations $\land,\lor,\lnot$ and they follow the rules of classical logic. Instead, Kolmogorovian probability considers only a special realization of a Boolean algebra, namely of the algebra of subsets of a set $\Omega$, with the set-theoretic operations $\cap,\cup,\bar{\quad}$ of intersection, union and complement. 

This leads to the natural question if there are abstract Boolean algebras which cannot be represented by the algebra of subsets of some space. The answer is given by Stone's representation theorem for Boolean algebras \cite{Stone}. It tells that every Boolean algebra is isomorphic to an algebra of subsets of some set $\Omega$, its Stone space.  The Stone space of a Boolean algebra is the set of homomorphisms of the algebra to the two-element Boolean algebra $\{0,1\}$. That means, each $\omega\in\Omega$ defines for each $A\in\mathcal{F}$ a truth value $A(\omega)\in \{0,1\}$. Every proposition $A\in\mathcal{F}$ is the equivalent to the subset $A \subset\Omega$ of propositions which are true for $A$, by $\{\omega\in\Omega|A(\omega)=1\}$. These subsets can be used to define a topology as well as the corresponding Borel algebra, so that every $A$ appears to be measurable.

Let's note here that the Stone space is a completely natural thing from point of view of plausible reasoning. An element of the Stone space $\omega\in\Omega$ fixes a truth value for every proposition, and, as a homomorphisms, does this in a consistent way, so that if $A$ and $B$ are assumed to be true, then $A\land B$ also has to be true. So, it is simply what one could describe as a consistent, conceivable, imaginable possibility, eventuality, story or scenario. So, the space $\Omega$ is a quite natural object, constructed in a straightforward way out of what is open and well-known, the set of consistent scenarios. 

This changes the interpretation of the space $\Lambda$, as we know it from usual discussions of Bell's theorem, where it considered to be a space of ``hidden parameters''. The space $\Omega\cong A\times \Lambda$ is, instead, constructed out of elements which are completely well-known -- the propositions of the field of discourse $\mathfrak{F}$ which is considered. Ok, the particular truth values fixed by the particular $\omega\in\Omega$ may remain hidden from observation, so in this sense the $\omega$ contains some hidden parameters. But the meaning of the particular parameter is well-defined and part of what is considered, and it is only the particular value which remains hidden. This completely contradicts the usual perceptions about hidden variables: The construction of the space of ``hidden variables'' is a trivial, straightforward operation, given the set of statements under consideration. 

Moreover, it also changes the interpretation of the role of random variables -- in particular of all observables -- being well-defined functions of $\omega$. One can naively think that behind this is hidden an assumption about determinism. It is not -- a random variable $f$ may be as fundamentally, inherently random as one likes, the proposition ``$f$ has the value $f_0$'' will be a valid proposition of our field of discourse, and therefore it is, for every $\omega\in\Omega$, either true or false, and in any case well-defined and fixed for this particular $\omega$. 

\section{Why logical independence is sufficient to ignore superdeterminism}

An important difference between the frequency interpretation of probability and the logic of plausible reasoning is the very different notion of independence. In the frequency interpretation, independence is a very nontrivial physical hypothesis -- the null hypothesis. It may appear false -- falsified by observation, which shows a correlation. The situation appears completely different in plausible reasoning. Here, the question is a completely different one -- what would be rational to think, given the available information? In a situation where we have no information which indicates any connection, it would be irrational to assume there exists one. Will $A$ increase or decrease the probability of $B$?  Without any information, we have no reason to prefer one of the two answers. The only rational answer is the one which makes no difference -- that $A$ is irrelevant for $B$, that both are independent of each other, $P(A\land B)=P(A)P(B)$. This approach, of course, does not prevent us from errors -- we may not have had the information required to recognize their correlation. But this is not a problem for plausible reasoning: Once the statistical data which suggest a dependence are known, they will be applied to modify the probabilities, in a process known as Bayesian updating. And despite the fact that the initial conclusion about the independence appeared to be wrong, this does not change the point that it was the only reasonable one, given the information which was available at this time.

This difference becomes essential for the consideration of superdeterminism -- the question if the general probability distribution $\rho(\omega)=\rho(a,\lambda)$ of \eqref{wishful-thinking} splits into $\rho(a)\rho(\lambda)$ of \eqref{split}. This is nothing but the usual situation -- we have the decisions of the experimenters, and in general, we have no information which would suggest that it is somehow correlated with anything relevant for the preparation or the execution of the experiment. And this lack of information is already sufficient, it is all we need in plausible reasoning to switch from \eqref{wishful-thinking} to \eqref{split}.

Let's note that this does not mean that superdeterminism is wrong. What we consider here is logic -- and nothing nontrivial about reality can follow from pure logic taken alone. And superdeterminism is, even if it is a completely unreasonable big conspiracy theory, a physical hypothesis. But we can ignore superdeterminism. Once we have switched to the logic of plausible reasoning, \eqref{split} no longer means that the physical hypothesis of superdeterminism has been somehow falsified, but that, given the information we actually have, it would be irrational to assume that superdeterminism is true. 

That means, logical independence allows us to obtain formula \eqref{split} out of  of \eqref{wishful-thinking} without the necessity to reject superdeterminism -- we can simply ignore it. Even if superdeterminism would be true, it would remain rational for us to use \eqref{split} instead of \eqref{wishful-thinking}.

This completes the first part of the paper, and gives the main result: Formula \eqref{no-wishful-thinking} appears to be a formula of the logic of plausible reasoning. It does not require any nontrivial physical hypothesis to justify this formula -- pure logic is sufficient. 

\section{Questioning the logic of plausible reasoning}

The most direct, straightforward possibility to preserve Einstein causality would be to question the logic of plausible reasoning. A rejection of logic sounds, of course, problematic. But that does not mean that there is no chance. First of all, one does not have to reject logic completely, it could be sufficient to weaken it. Moreover, there have been attempts, and, even if they were not really successful, they were also not completely unreasonable. Of course, a weakening of logic cannot be rejected in general, because we could not use logic to reject it. All what one can reject would be particular proposals to weaken logic, so that one could base a rejection on those parts of logic which remain untouched. Nonetheless, to look at the results of previous attempts to criticize logic appears interesting, and allows us to estimate what possible future attempts can give. 

\subsection{Relevance logic}

One such criticism against classical logic was ``relevance logic'' \cite{relevance-logic}. Many philosophers, in particular Hugh MacColl \cite{MacColl}, have claimed that some formulas of classical logic, like $p \to (q\to p)$, or $\lnot p \to (p\to q)$, or $(p \to q) \lor (q \to r)$ are counterintuitive if we interpret $\to$ as representing the concept of implication that we have before we learn classical logic. Relevance logicians claim that what is unsettling about these so-called paradoxes is that in each of them the antecedent seems irrelevant to the consequent. This has failed, not because relevance logic has failed to make proposals how to define implication more close to the common sense meaning of the word, but because for all purposes where implication has been used in logic before, it  was sufficient to use the $p \to q$ as simply a sign for $\lnot(p \land \lnot q)$ without some relevance beyond this. So, what has changed was not classical logic, but the interpretation of the meaning of the expression $p \to q$, removing ``relevance'' from its meaning. Given that a relevant implication  $p \to_{rel} q$ (however defined) also proves $p \to q$, logic was not even weakened by this reinterpretation. 

\subsection{Intuitionism}

Another classical criticism of logic were various variants of ``constructive mathematics'' \cite{constructive-mathematics}, in particular ``intuitionism''. They rejected  existence proofs of classical mathematics which have not given explicit constructions of the objects in question, and proposed different rules of logic, especially for the meaning of ``exists'', so that  ``there exists'' becomes close to ``we can construct''. This criticism was solved by the Gödel-Gentzen translation \cite{Goedel,Gentzen}, which in particular translated a classical  ``$(\exists x) A(x)$'' into  ``$\lnot ((\forall x) \lnot A(x))$''. As the result of this translation, every proof valid in classical logic became a valid proof in intuitionist logic too. So, the answer was quite similar: the mathematical apparatus of logic remained unchanged, and the criticism has only succeeded to modify the interpretation of ``exists'', distinguishing it from ``we can construct''. Similarly to the case above, classical logic was not weakened too -- an explicit construction can be used to prove existence too.  

So, the history of criticism of classical logic shows a clear pattern: The mathematical apparatus of logic remained unchanged. The only thing which has changed was the interpretation of the meaning of some mathematical expressions, away from the stronger common sense meaning toward a weaker interpretation which corresponds to the formulas of logic. While new mathematical constructions, which covered the common sense meaning in a more accurate way, have been proposed, they have not given much, and are now essentially abandoned. What has survived was the mathematical concept as used in pure logic, and the modified, weaker interpretation of the logical rule. 

\subsection{Frequentism vs. logic of plausible reasoning}

The difference between the frequentist interpretation and the Bayesian interpretation of probability theory -- the logic of plausible reasoning -- can be understood along similar lines. Jaynes \cite{Jaynes} sees the Bayesian interpretation in a classical tradition going back to Laplace, so that frequentism can be seen as a criticism of this tradition too. The key advantage of the logic of plausible reasoning was that it can be applied to a much wider range -- namely, it can be used to assign degrees of plausibility to theories and hypotheses, which cannot be interpreted as frequencies. On the other hand, similarly to the cases above, where the frequency interpretation is applicable, the resulting probabilities are the same.  

\subsection{Why this makes sense}

In all three cases, the result can be understood as being quite rational. Once the formulas of logic can be applied always and everywhere, it would not be reasonable to reject them. If there appears to be a contradiction between the common sense meaning of the phrase used to describe the logical rule and the formulas, the adequate reaction is, first, to find a more accurate description of the logical rule in common language, and, on the other hand, to find a new formalism which adequately describes the common sense meaning used before for the logical rule. Then, everybody is free to use what is more important for him. Given that the logical rule has a much wider range of applicability -- everywhere -- one can reasonably expect that the logical rule will be more widely used.

So, history and rationality seem to lead to similar conclusions: The formulas which are part of logic will survive. But it is possible that they have to be reinterpreted, and this reinterpretation has to be done in a careful way. One has to distinguish the purely logical notion -- usually weak, because applicable without any restriction -- from the common sense notion, which covers usually only archetypal applications instead of the most general one. The archetypal, common sense applications nonetheless remain valid applications, so that logic is not weakened by such a reinterpretation.  

\subsection{The situation with causality}

Regarding causality, the situation seems quite similar. With \eqref{no-wishful-thinking} as a logical formula, we have a sort of weak, purely logical notion of causal influence. It is defined by the function $o(a,\lambda)$, which always exists. This is different from the classical, common sense meaning of causality, which is considered to be a sort of physical hypothesis, which is part of some theories but not of others.  

The other thing we have observed in all three examples above, namely that if the stronger common sense notion is applicable, the weaker logical one is also valid, holds here too. In a theory with common sense causality, a dependence of the function $o(a,\lambda)$ on $a$ would be interpreted as a causal influence.

To summarize, there is no reason to hope that questioning the logic of plausible reasoning may allow to save Einstein causality. All one can hope for is a metaphysical reinterpretation of the meaning of causality.  

\section{The example of quantum theory}

It makes sense to consider this weaker notion of logical causality for quantum theory. At least in the minimal interpretation of quantum theory it is not a theory which makes statements about causality at all. The de Broglie-Bohm interpretation (which is essentially a different theory, which gives the quantum predictions only in a subset of quantum equilibrium states) has also been named ``causal interpretation'', with the obvious hint that all other interpretations are not causal. 

To consider the example of quantum theory is useful also because of other counterarguments. So it has been explicitly doubted that Kolmogorovian probability theory is sufficient to handle quantum theory, for example, by Khrennikov \cite{Khrennikov}:
\begin{quote}
Roughly speaking for any fixed experimental context a Kolmogorov space can be used. Probabilistic data collected in a fixed experiment in quantum (as well as classical) physics can be described by the conventional measure-theoretic model, but not data collected for a few incompatible observables. \cite{Khrennikov} p. vii
\end{quote}
Then, the Stone space we have used here is, in a general situation, a quite complex object. The problem is that to construct a particular $\omega\in\Omega$ in a situation where we have an infinite space of propositions $\mathcal{F}$ would require a similarly infinite amount of arbitrary choices. As a consequence, to prove it we have to use the axiom of choice\footnote{Technically, a slightly weaker assumption, the Boolean prime ideal theorem, is sufficient, but this would not change the point.}, which usually suggests that the construction is a quite artificial one. 

Fortunately, the axiom of choice is not as problematic as it is presented in popular culture. It is known today that this axiom does not lead to contradictions. If set theory without it is free of contradiction, it remains free of contradictions if we add the axiom of choice. Then, for the interesting applications this does not matter at all. In fact, whenever we consider a particular conceivable scenario, we try to fix only the relevant properties. This requires only a finite number of choices. And if we later have to consider some other property, which we have not fixed yet, we can fix it and continue. Stone's theorem is useful only because it gives us certainty that even in the purely theoretical case of an infinite number of propositions, which could force us into the necessity of an infinite number of such choices, would not lead us into contradictions. 

Nonetheless, given the importance of quantum theory, it seems useful to given an explicit and usable construction of this space for quantum theory.  

\subsection{Explicit construction for quantum theory}

For every particular quantum observation, as defined by an operator $\hat{F}$, quantum theory already defines a particular Kolmogorovian space with a probability distribution on it. The space is defined by the possible measurement results of the observable, say, $f$, together with the quantum state $\hat{\rho}$. The corresponding conceivable possibility would be the proposition ``the measurement of $F$ for the quantum state $\hat{\rho} \in \mathfrak{S}$ would give the result $f\in F$''. This gives us a space $\Lambda_F$ consisting of pairs $(\hat{\rho},f)\in \mathfrak{S}\times F \cong \Lambda_F$, with probability distribution $\rho(f)df$ defined by the eigenstates $f\in F$ of the operator $\hat{F}$ as $\langle \psi_f|\hat{\rho}|\psi_f\rangle$. 

Now let's consider an arbitrary set of incompatible measurements $\hat{F}_a, a \in A$, and combine them into a single big experiment, where we make some random choice $a\in A$ what to measure. For simplicity, we can restrict ourselves to a classical random device deciding about the $a\in A$. Let's denote the measurement results for different measurements as $f_a \in F_a$. What defines in this case the space $\Omega$?  

Given that it is conceivable that $\hat{F}_a$ is measured, we have to include the conceivable measurement results $f_a\in F_a$ for all $a\in A$. This gives us the product space $F_A\cong \prod_{a\in A} F_a$. The prepared state $\hat{\rho}\in\mathfrak{S}$ will be part too, which gives our resulting space $\Lambda \cong  \mathfrak{S}\times F_A$. The measurement which is actually made is, of course, also relevant, so that we obtain the space $\Omega \cong A \times \Lambda$ used in \eqref{no-wishful-thinking}. 

Let's define now the probability distribution on $\Omega$. For given $(a, \hat{\rho})$, quantum theory defines the probability distribution on the space $F_a$ of measurement results $\rho_a(f_a,\hat{\rho})$. On the product space, we can use 
\begin{equation}\label{quantum}
\rho(\lambda) = \rho(f_A,\hat{\rho})
	      =\prod_{a\in A}\rho_a(f_a,\hat{\rho}) \text{ on } \Lambda \cong  \prod_{a\in A} F_a \times \mathfrak{S}, 
\end{equation} 
which is a probability distribution of type $\rho(\lambda)d\lambda$ on $\Lambda$ we need. To show that this probability distribution is viable, all one needs is that it gives, for every fixed $a\in A$, the correct distribution on the space of the actual measurement results $F_a$, which is trivial by construction:  
\begin{equation}
\begin{split}
E(o|a) 	&= \int o(f_a)\rho(f_A,\hat{\rho}) d f_A\\
	&= \int o(f_a)\rho(f_a) df_a \prod_{a'\neq a}\rho_{a'}(f_{a'}) df_{a'} \\
	&= \int o(f_a)\rho(f_a) df_a.
\end{split}
\end{equation}

Given that it gives the correct predictions for all experiments allowed in quantum theory, there would be no need for any further justification. If this would have been a guess, so what? A positivist has certainly no reason to object. 

Nonetheless, this probability distribution can be derived from the principles of plausible reasoning. All we need to obtain \eqref{quantum} is, beyond the information provided by quantum theory about the probability distributions $\rho_a(f_a,\hat{\rho})$ of particular experiments, a basic principle of plausible reasoning, the principle of logical independence: If we have no information which indicates some dependence between two variables, then we have to assume that they are independent. 

We have to apply it two times. The first time when we have made the split $\rho(\omega)=\rho(a,\lambda) \to \rho(a)\rho(\lambda)$, assuming the independence of the free decisions of the experimenters what to measure, as discussed above. The second time when we have made the split of $\rho(f_A,\hat{\rho})$ into the product of the $\rho_a(f_a,\hat{\rho})$ in \eqref{quantum}. In both cases, we have, indeed, no information provided by quantum theory that there is a connection between the entities assumed as independent. 

\subsection{Details about causal dependencies would have to follow from hidden variable theories}

While the independence seems unproblematic for the free will decisions, the independence assumption made in \eqref{quantum} seems less plausible. Here one can reasonably expect that the future will give us additional information which will force us to reject the independence. What type of information could this be? We can give here a quite definitive answer, namely that it has to be a hidden variable theory. It has to provide some information which is not provided by quantum theory, that means, it is hidden information. What else is a hidden variable theory if not a theory which provides information not available in quantum theory?

Let's note here that the construction \eqref{quantum} in itself is not unknown. Kochen and Specker \cite{KochenSpecker} have given a variant of this construction, naming it ``somewhat trivial'' (which, indeed, it is) and ``artificial'' (it is not, but considering it as a ``hidden variable theory''  is indeed artificial, given that it does not contain any hidden information). The purpose for giving this construction was to ``to point out the insufficiency of condition (1) [essentially our \eqref{Kolmogorovian}] as a test for the adequacy of the solution of the [hidden variable] problem''. There is a reason to agree. That we can fulfill \eqref{Kolmogorovian} without adding any information to quantum theory means that fulfilling it does not ``solve'' any hidden variable problem. What it does is something different, namely to put an end on attacks on formula \eqref{Kolmogorovian}. One can hardly argue that this formula is inadequate and can no longer be applied if it can be obtained even for quantum theory as it is, based on a trivial construction. 

Kochen and Specker continued by proposing an additional condition for what would be an adequate solution of the hidden variable problem, today named non-contextuality, only to show that it cannot be fulfilled. While it is certainly important to find out properties which have to be violated by hidden variable theories, this is not a point of interest here. Our point is the opposite: The probability distribution  \eqref{quantum} may appear inadequate, because the independence assumption is too strong. This happens always if we have more information, information which indicates a dependence. No problem -- but what we can say is that another theory, which provides such information, will be a hidden variable theory in the most straightforward meaning of the word, simply because it provides -- it has to provide -- information which is hidden in quantum theory. 

A more detailed consideration of the question of hidden variable in quantum theory shows that quantum theory itself contains some elements which are not fixed by quantum theory, but depend on choices of the theoretician who describes a particular situation. Then, if we compare different choices of the theoretician, it appears that one description contains information which is hidden in the other one. 

One such theoretical decision is the classical-quantum split in the Copenhagen interpretation. Imagine one and the same experiment with different splits between the classical and the quantum part. Then, the description with the larger classical part contains information about the classical trajectory in the intermediate region, information which is hidden if the quantum part is larger. One split puts the mechanism which possibly kills the cat into the quantum part, one into the classical. Then, the second split contains the trajectory of the mechanism, which is hidden in the other split. 

Another situation where quantum theory differently interpreted appears to be a hidden variable theory is the possibility to describe the two measurements in Bell tests in different ways -- as a common measurement, or as two separate experiments, with two possible orderings in time. There exists either a dependence of $A(a,b,\lambda)$ on $b$ or of $B(a,b,\lambda)$ on $a$, as a consequence that Bell's inequality is violated. But if Alice measures first, then Bob's decision what to measure cannot influence the measurement result of Alice, so that $A(a,b,\lambda) = A(a,\lambda)$. Similarly, if Bob measures first, $B(a,b,\lambda) = B(b,\lambda)$.

These considerations already provide example of useful applications of the formalism. Once one recognizes that the space $\Lambda$ is a logical triviality, which does not indicate any nontrivial information about hidden variables, one can much easier identify those elements where hidden variables appear.

\section{Consequences: What remains from Einstein causality?}

Let's consider now the consequences of the main result. Einstein causality -- or, more accurate, any interpretation of Einstein causality which requires that $A(a,b,\lambda), B(a,b,\lambda)$ which appear in \eqref{Bell-realistic} have the form $A(a,\lambda), B(b,\lambda)$ -- is sufficient to derive the Bell inequalities, and therefore has to be rejected. What could be a replacement? One radical solution would be, in principle, to give up causality completely. But, as we will see, there is no necessity for this.  

\subsection{Weakening Einstein causality: Signal causality}

One straightforward possibility, which seems quite popular even now, is to give up the usual, strong version of Einstein causality, and to go back to a weaker notion of causality -- weak or signal causality. With weak causality, it is not sufficient for a causal connection that some function $o(a,\lambda)$ depends on some variable $a$, but, first, we care only about observables $o(a,\lambda)$, only if they depend on free decisions of experimenters $a$, and we require not only that the function $o(a,\lambda)$ itself depends on $a$, but if the expectation value $E(o|a) = \int o(a,\lambda)\rho(\lambda)d \lambda$ depends on $a$. 

While such a restriction, with its restriction to observable effects, is quite plausible from point of view of positivism, positivism is no longer a reasonable candidate for a philosophy of science. In Popper's critical rationalism, scientific theories are hypotheses, thus, can and will contain hypotheses which are not immediately open to empirical tests. They have to make testable empirical predictions, the more the better, but if some hypothesis about unobservable elements of reality leads, however implicit, to some additional testable predictions, this is sufficient to justify the hypothesis. 

Such a weak notion of causality makes, of course, sense as a useful information about restrictions for human devices, preventing hopeless affords similar to perpetuum mobile construction. But as some fundamental principle it seems useless. The point is that now we know that for every theory there will exist some description in terms of $o(a,\lambda)$ and $E(o|a) = \int o(a,\lambda)\rho(\lambda)d \lambda$, so that the information that $E(o|a)$ does not depend on $a$ either follows from $o(a,\lambda)$ not depending on $a$, which would mean strong causality, or it is a consequence of the impossibility to fix in our preparations sufficiently precise distributions $\rho(\lambda)d\lambda$, thus, it would be only a side effect of restricted human abilities with no fundamental importance. 

So, even if one accepts weak causality as an interesting information, one would not consider it as something fundamental. It is derived from the fundamental, strong notion of causality defined by the functions $o(a,\lambda)$ together with restrictions of our ability to prepare initial states. 

\subsection{Classical causality in the Lorentz ether}

If one wants to preserve for a fundamental causality at least the non-existence of causal loops (as necessary to avoid the grandfather paradox), one has no choice but to accept a hidden preferred frame.

Indeed, assume we can, for any pair of events, observe violations of the Bell inequalities. Let's fix the event $e_a$ when Alice makes the measurement and try to identify an equal time event on Bob's world-line $e_b(t)$. We assume that we have moments $t_-$ long before the experiment and $t_+$ long after so that $e_b(t_-)\to e_a \to e_b(t_+)$. Then all we have to do is to use Dedekind cuts to construct a sequence  $(t^n_-,t^n_+)$ with $e_b(t^n_-)\to e_a \to e_b(t^n_+)$ and $|t^n_+-t^n_-|=2^{-n}|t_+-t_-|$. We observe a violation of the Bell inequalities for $t^{i+1}=\frac12 (t^i_+ + t^i_-)$. Based on strong causality, we conclude that either $e_a \to e_b(t^{i+1})$ or $e_b(t^{i+1})\to e_a$, so that we can put $t^{i+1}$ either in place of $t^{i+1}_+$ or $t^{i+1}_-$, reusing the other $t^i_{\pm}$ as $t^{i+1}_{\pm}$. This gives $e_b(t^{i+1}_-)\to e_a \to e_b(t^{i+1}_+)$. This sequence gives a limit $t_*$ with the property that for every $\varepsilon>0$ we have $e_b(t_*-\varepsilon)\to e_a \to e_b(t_*+\varepsilon)$. The $e_b(t_*)$ defines the moment contemporary to $e_a$, so that we obtain in this way a global contemporaneity. 

Given that we can only observe a violation of the Bell inequalities, but cannot find out what was the direction of causal influence which has caused it, the resulting preferred foliation remains hidden from observation. But is a hidden preferred frame compatible with modern physics? 

In the domain of special relativity, this would require nothing but a return to the Lorentz interpretation, where a preferred rest frame of the Lorentz ether exists, and relativistic effects are distortions of clocks and rulers caused by velocity against a static homogeneous ether. A generalization of the Lorentz ether into the domain of relativistic gravity which gives, in a natural limit, the Einstein equations of GR in harmonic coordinates, is also known \cite{gle}. This generalized Lorentz ether is no longer static and homogeneous, and its density and stress tensor also influences clocks and rulers. There exists also a proposal for an interpretation of the standard model of particle physics compatible with the Lorentz ether \cite{clm}. This model gives the SM gauge group and the three generations of fermions, together with some additional massive scalar fields. So, compatibility of the Lorentz ether with modern physics is unproblematic. The classical notion of causality can also survive in quantum theory, as, for example, in realistic and causal interpretations of quantum theory like the de Broglie-Bohm interpretation.  Moreover, given that in the Lorentz ether causality is connected with the fixed Newtonian background of absolute space and time, to preserve causality during quantization of the Lorentz ether is unproblematic too -- the background remains fixed, so that causality remains fixed too, and is not endangered by uncertainty of the metric. 

\section{Conclusion}

We have shown that a basic formula in the proof of Bell's inequality, \eqref{Bell-realistic}, is a formula following from the logic of plausible reasoning applied to the situation of the experiment considered in Bell's paper, without any further physical assumptions like realism or causality. As a consequence, all what is necessary to prove Bell's theorem is Einstein causality. That means, given that the Bell inequalities are violated for space-like separated pairs of measurements, Einstein causality has been empirically falsified. 

None of the loopholes usually discussed -- rejection of realism and causality, or superdeterminism -- are left. The only possibility to preserve Einstein causality would be to reject the logic of plausible reasoning. We have considered earlier attempts to criticize classical logic, and came to the conclusion that all one can hope for is a reinterpretation of the meaning of the formulas of logic, away from the common sense meaning toward a weaker, more general, purely logical interpretation. Such a reinterpretation of the meaning of causality can be found here too. 

We have illustrated this considering quantum theory, where we have identified a rather weak notion of causality. We have found that any strengthening of this weak notion would require some hidden variables. 

Last but not least, we have considered the possibilities to weaken Einstein causality. One possibility would be a notion of weak Einstein causality based on signal causality. Such a notion of causality cannot have any fundamental importance, because the impossibility to send signals faster than light is only a consequence of uncertainty relations preventing the preparation of sufficiently certain states. If one wants to have a fundamental notion of causality without causal loops, one has to accept the introduction of a hidden preferred frame. Fortunately, nothing in modern science is in conflict with such a hidden preferred frame, theories based on such a preferred frame exist as for relativistic gravity as for the standard model of particle physics.

\end{document}